\documentclass[prl,twocolumn]{revtex4}
\usepackage{graphicx}
\usepackage{color}

\begin{document}
\title{Quarkonium Production and Medium Effects in High Energy Nuclear Collisions}
\author{Kai Zhou$^1$}
\author{Nu Xu$^{2,3}$}
\author{Pengfei Zhuang$^1$}

\affiliation{ $^1$Department of Physics, Tsinghua University, Beijing, China}
\affiliation{ $^2$College of Physical Science and Technology, Central China Normal University, Wuhan, China}
\affiliation{ $^3$Nuclear Science Division, Lawrence Berkeley National Laboratory, Berkeley, CA 94720, USA}
\date{\today}

\begin{abstract}
Color screening and regeneration are both hot medium effects on
quarkonium production in high energy nuclear collisions. However,
they affect in an opposite way the finally observed quarkonium
spectra. Due to the competition of the two dynamical effects, the
ratio of the integrated quarkonium yield between nuclear and
elementary nucleon collisions loses its sensitivity. Once the
information of quarkonium transverse motion is included, on the
other hand, the ratio of averaged transverse momentum square reveals
the nature of the QCD medium created in high energy nuclear
collisions.
\end{abstract}

\maketitle

The existence of a deconfined partonic phase has been well
established in $\sqrt {s_{NN}}$ = 200 GeV Au+Au collisions at
RHIC~\cite{rhicexp1,rhicexp2,rhicexp3,rhicexp4,rhicthe} as well as in $\sqrt {s_{NN}}$ = 2.76 TeV Pb+Pb
collisions at LHC~\cite{lhcexp1,lhcexp2,lhcexp3}. The study of high energy
nuclear collisions has entered a new era where current physics
programs at both RHIC and LHC focus on the properties of the new
form of matter and how the hot matter undergoes the transition from
the quark gluon plasma to the hadronic gas. Due to color screening,
quarkonium suppression in nuclear collisions in comparison with the
elementary p+p collisions has been proposed to "provide an
unambiguous signature of quark gluon plasma
formation"~\cite{satz86}. Regeneration~\cite{regeneration1,regeneration2,regeneration3,regeneration4}, that is quarkonium
production in the hot medium via coalescence process, on the other
hand, ruins the simplistic picture.

For the final quarkonium distributions emerging from high energy nuclear
collisions there are several genuine contributors, namely, (1)
initial production via pQCD process; (2) cold nuclear matter effects
before the formation of the hot medium including the shadowing
effect~\cite{sha,shadowing,shadowing2}, Cronin effect~\cite{cronin1,cronin2,cronin5}
and nuclear absorption~\cite{abs}; and (3) Debye screening and regeneration.
While screening and regeneration are both hot nuclear matter effects,
they affect the quarkonium production in an opposite way.
The yield is suppressed by the former but enhanced by the latter.

While the regeneration is negligible and
the initial production is dominant for low energy
collisions~\cite{regeneration4,karsch06}, both initial production and regeneration
are important for nuclear collisions at high energies. The nuclear
modification factor $R_{AA}=N_{AA}/\left(n_{bin} N_{pp}\right)$ is used
commonly to study the nuclear matter effects on the quarkonium
production, where $n_{bin}$ is the number of binary collisions,
$N _{pp}$ and $N_{AA}$ are the
integrated quarkonium yields in p+p and A+A collisions,
respectively. This quantity has been measured at all collision
energies at SPS, RHIC and LHC. Starting at about unity in peripheral
collisions, the $R_{AA}$ decreases towards more central collisions.
Although the collision dynamics from SPS to LHC energy is dramatically different, there is no clear energy dependence in the integrated
$R_{AA}$~\cite{digal12}.

In order to study the medium effects in high energy nuclear
collisions, one clearly needs a new observable with higher
sensitivity to the underlying collision dynamics. We start with the
introduction to the model we used in this analysis. To extract
medium information from the quarkonium motion, both the medium and
the quarkonia created in high energy nuclear collisions must be
treated dynamically. Since a quarkonium is heavy, its phase space
distribution $f_\Psi({\bf p},{\bf x},t)$ is governed by a
Boltzmann-type transport equation~\cite{tsinghua1,tsinghua2},
\begin{equation}
{\partial f_\Psi\over \partial t} +{\bf v}_\Psi\cdot{\bf \nabla}f_\Psi=-\alpha_\Psi f_\Psi +\beta_\Psi,
\label{trans}
\end{equation}
where both the initially produced and regenerated quarkonia are
taken into account through the initial distribution $f_\Psi$ at the
medium formation time $\tau_0$ and the gain term $\beta_\Psi({\bf
p},{\bf x},t)$, and ${\bf v}_\Psi={\bf p}_\Psi/E_\Psi$ is the
quarkonium velocity. The reduction of quarkonium due to the Debye
screening is described by the lose term $\alpha_\Psi({\bf p},{\bf
x},t)$. The cold nuclear matter effects change the initial
quarkonium distribution and heavy quark distribution at $\tau_0$.
The interaction between the quarkonia and the medium is reflected in
the lose and gain terms and depends on the local temperature $T({\bf
x},t)$ and velocity $u_\mu({\bf x},t)$ of the medium which are
controlled by the hydrodynamics~\cite{heinz,hirano}.

\begin{figure}[htbp]
\includegraphics[width=0.43\textwidth]{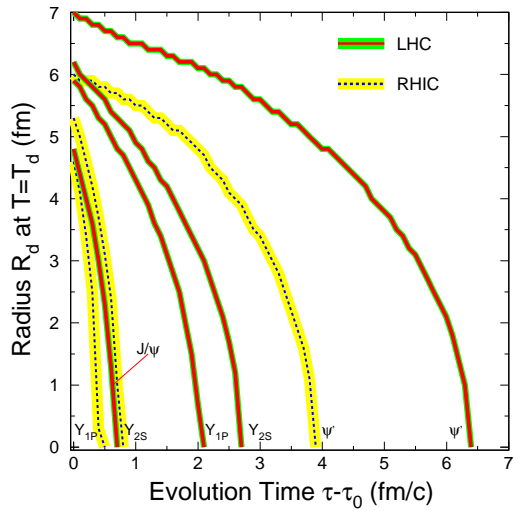}
\caption{(color online) Averaged quarkonium dissociation radius
$R_d$ in transverse plane as a function of the medium evolution time
$\tau$ in central Au+Au collisions at RHIC (dashed lines) and Pb+Pb
collisions at LHC (solid lines).  }
\label{fig1}
\end{figure}

The Debye radius depends strongly on the properties of the medium
such as temperature as well as the strength of the binding potential
of the quarkonium states. By solving the Schr\"odinger equation for
the bound state of a pair of heavy quarks, one obtains the
quarkonium dissociation temperature $T_d$~\cite{satz06}. From the
hydrodynamic solution $T({\bf x},t)$ of the fireball created in
heavy ion collisions, one can extract the dissociation radius
$R_d(\tau,y)$ determined by
\begin{equation}
\label{rd}
T(R_d,y,\tau)=T_d
\end{equation}
in the transverse plane as a function of the proper time $\tau$ and
rapidity $y$. Fig.\ref{fig1} shows the mid-rapidity $R_d(\tau)$ for
central Au+Au collisions at RHIC (dotted lines) and Pb+Pb collisions
at LHC (solid lines). The quarkonium state is destroyed if it is in
the region $r<R_d$. The dissociation temperatures used in
Fig.\ref{fig1} are taken as $T_d/T_c =  4,\ 1.8,\ 1.6,\ 2.3$ and 1.1
for $\Upsilon_{1S}$, $\Upsilon_{1P}$, $\Upsilon_{2S}$, $J/\psi$ and
$\psi' (\chi_c)$~\cite{satz06,guo}, respectively, scaled by the
critical temperature of deconfined phase transition $T_c = 165$
MeV~\cite{karsch00}. The temperature of the medium created at LHC is
expected to be much hotter than that at RHIC. As a result one finds
$R_d \rm{(LHC)}
> R_d \rm{(RHIC)}$ for all quarkonium states. Since $T_d$ is chosen
so high for $\Upsilon(1S)$ and $J/\psi$, the directly produced
$\Upsilon(1S)$s are not destroyed even at LHC and some of $J/\psi$s
can survive at RHIC.

\begin{figure}[htbp]
\includegraphics[width=0.43\textwidth]{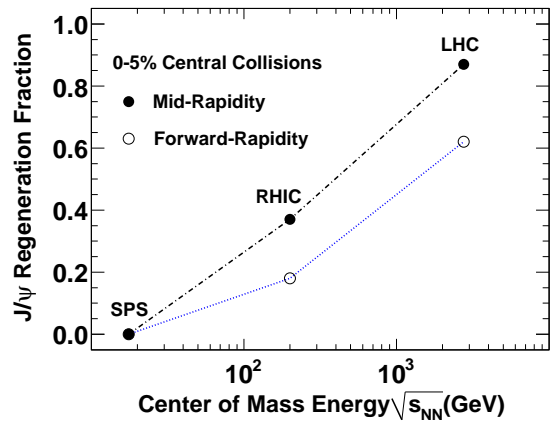}
\caption{(color online)  $J/\psi$ regeneration fraction in
central Au+Au collisions at RHIC and Pb+Pb collisions at SPS and LHC.
Results for mid- and forward-rapidity regions are shown as filled
and open circles, respectively. }
\label{fig2}
\end{figure}

In the most violent collisions, quark-gluon plasma is formed and the
emerging hadrons are created via coalescence process. For quarkonia
it is the regeneration that dominates the production at high
energies. Different from the light hadrons which are produced at the
hadronization surface determined by $T({\bf x},t)=T_c$, the
quarkonium regeneration happens continuously in the parton phase.
The fraction of the regenerated $J/\psi$s,
\begin{equation}
\label{fraction}
g_{AA}={N_{AA}^{reg}\over N_{AA}},
\end{equation}
calculated from the transport equation (\ref{trans}), is shown in
Fig.~\ref{fig2} as a function of the collision energy, where
$N_{AA}^{reg}$ is the integrated $J/\psi$ yield from the
regeneration. As expected, the fraction depends on heavy quark
density and, in turn, on the collision energy. At SPS, initially produced heavy quarks are few, there is almost
no regeneration. At RHIC, both initial production and regeneration of charmonia
play important role. At LHC, the $J/\psi$ production is dominated
by regeneration. The lower fraction in the forward rapidity merely
reflects the rapidity distribution of heavy quarks.

In heavy ion collisions, transverse motion is developed during the
dynamical evolution of the system. The microscopically high particle
density and multiple scatterings are essential for the finally
observed transverse momentum distributions. The distributions are
therefore sensitive to the medium properties, like the equation of
state. The study on the transverse motion has been well documented
in light quark sectors at all
energies~\cite{rhicexp1,rhicexp2,rhicexp3,rhicexp4,hecke98}. In
order to understand the quarkonium production and suppression
mechanisms and extract the properties of the medium, we propose to
construct a new ratio of the second moment of the transverse
momentum distribution. The ratio $r_{AA}$ is defined as
\begin{equation}
r_{AA}=\frac{\langle p_t^2 \rangle_{AA}}{\langle p_t^2
\rangle_{pp}}.
\end{equation}
The reason to choose $\langle p_t^2\rangle$ instead of $\langle
p_t\rangle$ is that we are interested in the medium induced change
in the shape of the transverse momentum distribution not only
its average value.

\begin{figure}[h]
\includegraphics[width=0.52\textwidth]{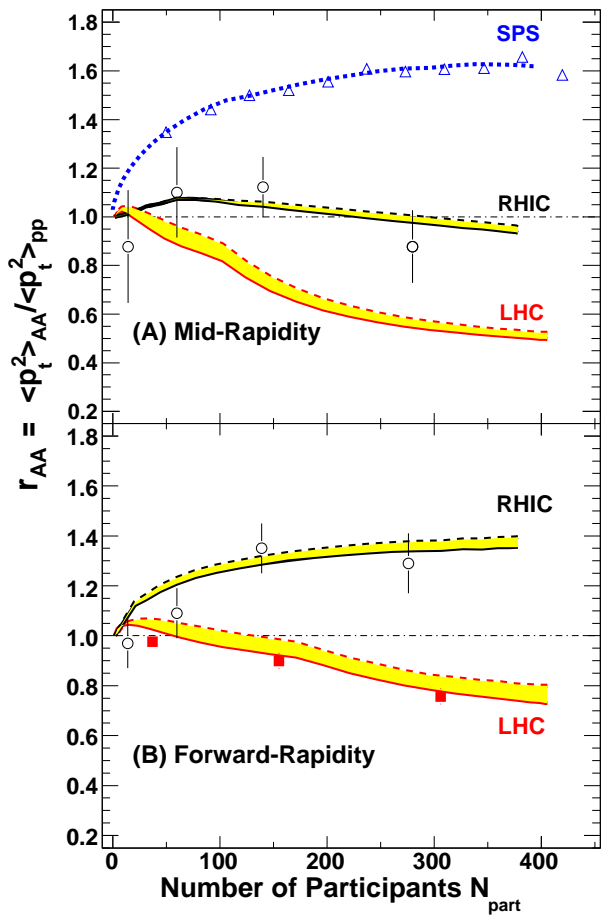}
\caption{(color online)  $J/\psi$ $r_{AA}=\langle
p_t^2\rangle_{AA}/\langle p_t^2\rangle_{pp}$ as a function of the
number of participants $N_{part}$ for Pb+Pb collisions at SPS energy
$\sqrt {s_{NN}}=17.3$ GeV (open triangles)~\cite{spsptraa}, Au+Au
collisions at RHIC energy $\sqrt{s_{NN}}=200$ GeV (open
circles)~\cite{rhic2007} and Pb+Pb collisions at LHC energy
$\sqrt{s_{NN}}=2.76$ TeV~(filled circles)\cite{aliceptraa}. The upper and
lower plots are results for mid-rapidity ($|y|<1.0$ at SPS, $<0.35$ at RHIC and $<2.4$ at LHC) and forward-rapidity ($1.2<y<2.2$ at RHIC and $2.5<y<4$ at LHC),
respectively.  The solid and dashed lines indicate the model results
with and without considering the shadowing effect~\cite{shadowing}.
} \label{fig3}
\end{figure}

Fig.\ref{fig3} shows the ratio $r_{AA}$ as a function of the number
of participant nucleons $N_{part}$ for collisions at
SPS~\cite{spsptraa}, RHIC~\cite{rhic2007} and LHC~\cite{aliceptraa}.
Starting from the very small number of participants, there is a clear rise in $r_{AA}$ at all
collision energies. It is believed due to the initial gluon multiple
scatterings before the quarkonium formation~\cite{cronin1,cronin2},
and the enhancement seems more important for collisions at lower energies.

The energy dependence of $r_{AA}$ clearly reflects the underlying
$J/\psi$ production and suppression mechanisms in high energy
nuclear collisions. At lower collision energy where the charm cross
section is small and the regeneration is negligible, almost all of
the observed $J/\psi$s are from direct production at the initial
impact and the decay contribution from the excited states $\psi'$
and $\chi_c$. In this case, the initial state gluon scattering,
$i.e.$ the Cronin effect~\cite{cronin1,cronin2} is dominant and
tends to increase the transverse momentum of the finally observed
$J/\psi$s. Since the Cronin effect is proportional to the gluon
traveling length in the nuclei~\cite{cronin3,cronin4}, $r_{AA}$ is
always above unity and increases monotonically versus collision
centrality, see upper panel of Fig.~\ref{fig3}. In the truly high
energy nuclear collisions, on the other hand, charm quarks are
copiously produced and the regeneration for charmonia, $\it{i.e.}$
coalescence of heavy quark and anti-quark pair inside the quark
gluon plasma, can be significant. Although these initially produced
heavy quarks carry high transverse momentum, they lose energy
(momentum) when passing through the
medium~\cite{quench1,quench2,quench3,quench4}. Therefore, as a
consequence of the increasing regeneration fraction with colliding
energy, see Fig.\ref{fig2}, the competition between the initial
production which controls high $p_t$ charmonium production and the
regeneration which inherits the low momentum of thermalized heavy
quarks leads to the decrease of the values of the ratio $r_{AA}$
from SPS to LHC. As shown in the upper panel of Fig.\ref{fig3}, the
predicted $r_{AA}$ at mid rapidity for heavy ion collisions at LHC
is below unity and decreases toward more central collisions. The
prediction has been confirmed by the experimental results at
somewhat forward rapidity window as shown in the lower panel of
Fig.\ref{fig3}~\cite{aliceptraa}. At RHIC energy, the competition
between the initial gluon scattering and the final stage
regeneration leads to a weak centrality dependence for the
mid-rapidity $r_{AA}$. On the other hand, at the forward-rapidity,
due to the lower heavy quark production cross section, the
regeneration gives its way to the initial gluon scattering, and
$r_{AA}$, shown in the lower panel of Fig.\ref{fig3}, becomes higher
than unity and increases as a function of $N_{part}$. Since the
heavy quark production cross section is large at LHC, even at the
forward-rapidity, the $r_{AA}$ remains lower than unity.
\begin{figure}[h]
\includegraphics[width=0.42\textwidth]{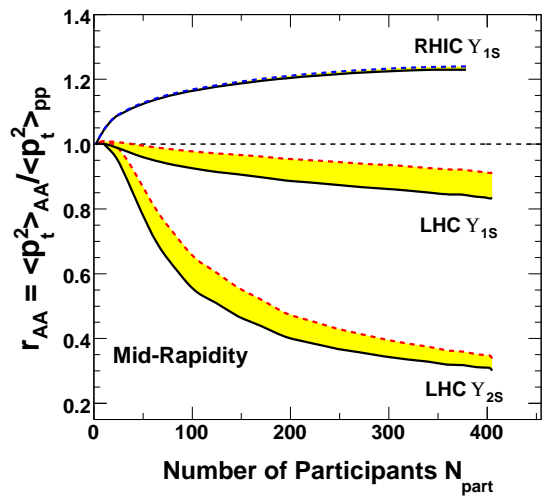}
\caption{(color online)  Prediction on $\Upsilon$ $r_{AA}=\langle
p_t^2\rangle_{AA}/\langle p_t^2\rangle_{pp}$ as a function of the
number of participants $N_{part}$ for Au+Au collisions at RHIC
energy $\sqrt{s_{NN}}=200$ GeV ($|y|<0.35$) and Pb+Pb collisions at LHC energy
$\sqrt{s_{NN}}=2.76$ TeV ($|y|<2.4$). The solid and dashed lines indicate the
calculations with and without considering the shadowing
effect~\cite{shadowing}.} \label{fig4}
\end{figure}

One may ask oneself if the behavior of $r_{AA}$ is caused by the
initial shadowing effect which changes the parton distributions in
nuclei. Different from the integrated yield, the averaged transverse
momentum is a normalized quantity, therefore the shadowing induced
change in the parton distribution is minimized in the final
quarkonium distribution. As shown in Fig.\ref{fig3}, the small
difference between the solid and dashed lines, the hatched band, is
the results of shadowing effect~\cite{shadowing}. At RHIC energy the
band becomes very narrow.

Comparing to the yield $R_{AA}$, the $p_t$ $r_{AA}$ lends additional
sensitivity to the transverse dynamics. It is known that transverse
motion plus local properties of the medium are extremely important
for hadron productions, including quarkonia, in heavy ion
collisions. Since the fireballs created in heavy ion collisions at
RHIC and LHC energies are so hot, the quarkonium suppression and
regeneration in the hadron phase at later stage is assumed to be
neglected and all of the effects shown in Fig.\ref{fig3} are
originated from the partonic phase.

Except the mass difference, our results for $J/\psi$ should also
work qualitatively for heavier $\Upsilon$. The $\Upsilon$ $r_{AA}$
is shown in Fig.\ref{fig4} at RHIC and LHC energies. Since the
degree of both suppression and regeneration for $J/\psi$ at lower
energies is similar to that for $\Upsilon_{1S}$ at higher energies,
$J/\psi$ at SPS (RHIC) behaves like $\Upsilon_{1S}$ at RHIC (LHC), see the comparison between
Fig.\ref{fig4} and the upper panel of Fig.\ref{fig3}. For the
excited state $\Upsilon_{2S}$, its dissociation temperature is much
lower than the fireball temperature at LHC, therefore, the initial
production is significantly suppressed and the dominant regeneration
leads to a very small $p_t$ $r_{AA}$. These predictions will be
checked by the future experimental data from RHIC and LHC.

To summarize we argue that in high energy nuclear collisions the
quarkonium transverse momentum distribution can be used as a sensitively probe for studying the properties of
the quark gluon plasma. The Debye
screening and regeneration are both medium induced effects but affect the quarkonium production in an opposite way. When Debey screening is
important, initially produced quarkonia are destroyed and the
finally observed quarkonia are dominantly from the regeneration in
the hot medium. We promote the ratio $r_{AA}=\langle
p_t^2\rangle_{AA}/\langle p_t^2\rangle_{pp}$ for quarkonia as a
sensitive probe for studying the nature of the hot medium. Its
dramatic energy dependence can be used to distinguish between the
hot mediums at SPS, RHIC and LHC.

\appendix {\bf Acknowledgement}: We thank Dmitri Kharzeev and Ralf Rapp for helpful discussions.
The work is supported by the NSFC, MOST and DOE grant Nos. 11079024,
11221504, 11335005, 2013CB922000 and DE-AC03-76SF00098.


\end{document}